\documentclass[a4paper,11pt]{article}
\usepackage{jinstpub} 
\usepackage{lineno}
\usepackage{float}
\usepackage{graphicx}
\usepackage{float}
\usepackage{subcaption}
\usepackage{caption}

\title{\boldmath Progress in Chromatic Calorimetry Concept: Improved Techniques for Energy Resolution and Particle Discrimination}

\author{Devanshi Arora\,$^{1,2,*}$, Matteo Salomoni\,$^{1,3}$,Yacine Haddad\,$^{4}$,Vojtech Zabloudil\,$^{1,5}$,Michael Doser\,$^{1}$, Masaki Owari\,$^{2}$, Etiennette Auffray\,$^{1}$}
\affiliation{$^{1}$European Organization for Nuclear Research (CERN), Geneva, Switzerland\\
$^{2}$Shizuoka University, 3 Chome-5-1 Johoku, Chuo Ward, Hamamatsu, Shizuoka 432-8011, Japan\\
$^{3}$University of Milano-Bicocca. Piazza dell'Ateneo Nuovo, 1, 20126 Milan, Italy\\
$^{4}$ Northeastern University, Boston, USA\\
$^{5}$ Czech Technical University in Prague, Prague, Czech Republic\\
$^{*}$Corresponding author}

\emailAdd{devanshi.arora@cern.ch}

\abstract{This study introduces chromatic calorimetry, a novel particle detection method that uses strategically layered scintillators with different emission wavelengths. This approach aims to enhance energy measurement by capturing particle interactions at different shower depths through wavelength-based discrimination. Our experimental validation of this novel method utilizes an arrangement of scintillator materials to improve energy resolution and particle identification. The stacking arrangement of scintillators is organized by their emission wavelengths to track the development of electromagnetic and hadronic showers. By testing electrons and pions with energies up to 100 GeV, the setup demonstrated better discrimination and provided detailed measurements of shower profiles. The results indicate that our experimental validation significantly aims to enhance particle identification and energy resolution, highlighting its potential value in high-energy particle detection. Future work will explore the integration of Quantum Dots (QD) technology to advance these capabilities in chromatic calorimetry further.}


\begin{document}
\maketitle
\flushbottom

\section{Introduction}
In particle physics, calorimeters are essential instruments used to measure the energy of particles as they interact with detector materials. They play a critical role in particle accelerator experiments, where understanding fundamental particle properties is vital. Homogeneous calorimeters use a single material for energy absorption and signal detection, allowing them to achieve excellent energy resolution \cite{RevModPhys.88.015003}. Common materials, such as lead tungstate (PbWO4) are favored for their scintillation properties and high density, with PbWO4 being notably employed in the electromagnetic calorimeters of both the CMS and ALICE experiments at LHC, CERN \cite{CERN-LHCC-97-031, alice1998technical}. However, the homogenous calorimeters often lack longitudinal segmentation, which makes it difficult to capture detailed profiles of particle showers as they develop. Understanding the longitudinal development of these showers is crucial for accurately identifying particles and studying their interactions. When a high-energy particle interacts with matter, it produces secondary particles that create a cascade known as a shower. An ideal homogeneous calorimeter accurately measures and reconstructs the shower, allowing it to determine the original energy and type of the incoming particle. However, without longitudinal segmentation, discerning subtle differences in shower profiles becomes challenging. This highlights the need for continued advancements in calorimeter technology to improve particle identification and enhance our understanding of fundamental interactions in high-energy physics experiments \cite{Apostolakis_2023}. To overcome the limitations of longitudinal segmentation, chromatic calorimetry is introduced by M.Doser et al. \cite{doser2022}, which utilizes varying scintillator materials to differentiate between different wavelengths of emitted light, offering a potential solution to the limitations of homogeneous calorimeters. This concept takes advantage of the tunability and narrow emission bandwidth (approximately 20 nm) of various materials, such as quantum dots, and quantum wells \cite{Zhang_2021, yuan2018}. By layering scintillator materials with different emission wavelengths, we aim to enhance the measurement of energy deposited by particles during their interactions. Our previous test beam study validates this proposed concept using inorganic scintillators strategically stacked by decreasing emission wavelength. Using a 100 GeV beam, we achieved analytical discrimination between electron and pion events and the reconstructed energy signals also reveal valuable separability between hadronic and electromagnetic interactions at higher energies, enhancing our ability to differentiate between these events \cite{arora2024enhancingenergyresolutionparticle}.
To enhance energy resolution, improve particle identification, and better track the development of electromagnetic and hadronic showers, this paper presents a modified iteration of the previously conducted chromatic calorimetry experiment with a new arrangement of scintillator materials. We conducted tests using electrons and pions with energies up to 100 GeV, demonstrating improved discrimination and providing detailed measurements of shower profiles. The results indicate that this modified design significantly enhances particle identification and energy resolution, underscoring its potential value in high-energy particle detection. However, it is important to note that our current experimental iterations do not incorporate quantum dots. Instead, we focus on optimizing the arrangement of scintillator materials to enhance particle identification and energy resolution in our chromatic calorimetry setup. Looking ahead, the ultimate goal of this research is to develop a quantum dot-based chromatic calorimeter that utilizes the unique properties of quantum dots for even greater performance in future applications. This will allow for more precise measurements and better integration within detection systems.

In this paper, we investigate an experiment to validate the chromatic calorimetry concept which was initially introduced by M. Doser et al. \cite{doser2022}. We have designed a crystal stack with scintillators having different emission wavelengths, which was subjected to to high energy beam of 100 GeV.

\section{Materials and methods}
\subsection{Crystal stack description}
\begin{figure}[H]
\centering
\begin{subfigure}{\textwidth}
    \centering
    \includegraphics[width=0.8\textwidth]{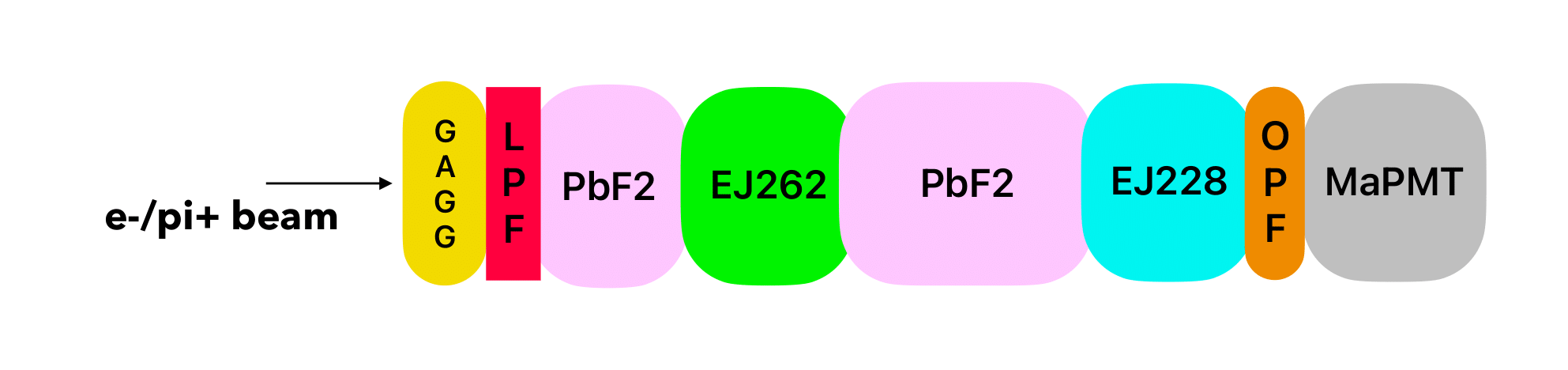}
\end{subfigure}

\vspace{1em}

\begin{subfigure}{0.48\textwidth}
    \centering
    \includegraphics[width=\textwidth]{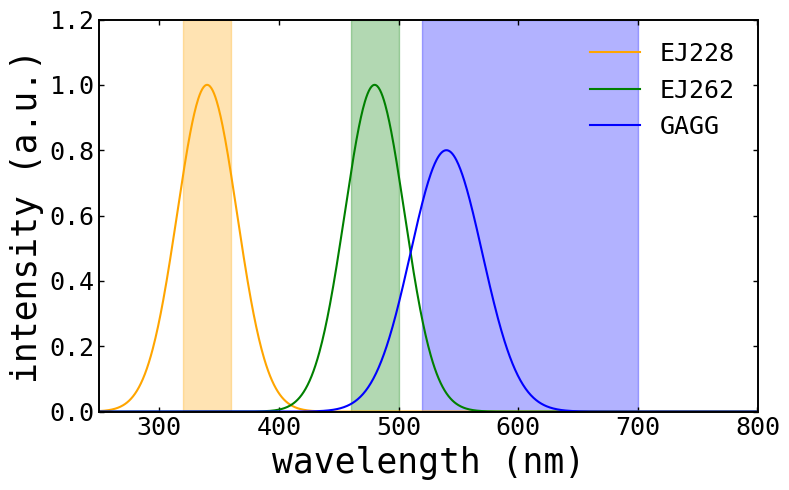}
\end{subfigure}
\hfill
\begin{subfigure}{0.5\textwidth}
    \centering
    \includegraphics[width=\textwidth]{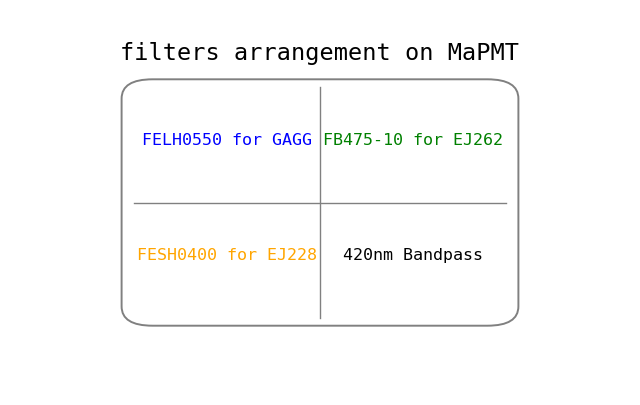}
\end{subfigure}

\caption{Chromatic calorimeter module schematic: The module comprises scintillators, one Thorlabs long pass FELH0550 (LPF) placed after GAGG, and read out by a MaPMT with four channels. Four different OPF are used, where for three channels of the MaPMT: Thorlabs long pass FELH0550 for GAGG, short pass FESH0400 for EJ228, and bandpass FB475-10 for EJ262, and the fourth channel a 420 nm bandpass filter (BPF), which was chosen to avoid interference with the emission characteristics of the other scintillator materials \cite{thorlabfilters}.}
\label{fig-1}
\end{figure}

The crystal stack was constructed using the following scintillators, shown in Fig.\ref{fig-1} (with the last dimension along the longitudinal shower propagation): 2x2x2 cm$^3$ gadolinium aluminum gallium garnet (GAGG, 540 nm peak emission), 2x2x5 cm$^3$ and 2x2x12 cm$^3$ lead fluoride (PbF$_2$, Cherenkov emitter), 2x2x2 cm$^3$ EJ262 (plastic scintillator, 481 nm peak emission), and 2x2x2 cm$^3$ EJ228 (plastic scintillator, 391 nm peak emission). These plastic scintillators were purchased from Scionix Holland B.V. \cite{scionixnl}. This arrangement was designed to capture scintillation light from three scintillators: at the beginning of the shower (in GAGG), at the shower maximum (in EJ262), and the end of the shower (in EJ228). PbF$_2$ crystals, chosen for their high density and transparency to the emission wavelengths of the other scintillators, were incorporated within the stack to facilitate shower development between GAGG, EJ262, and EJ228. PbF$_2$ crystals have short Radiation Length, Moliere Radius, high stopping power, and relatively low Cherenkov emission\cite{ACHENBACH1998357}.

The order of the scintillators along the shower propagation was carefully optimized to minimize re-absorption effects and ensure efficient photon transmission through the stack. A long-pass filter (LPF) FELH0550, was added after GAGG to ensure that light emitted from GAGG is well-separated from other scintillators. The experimental setup was completed with a Hamamatsu Multi-anode Photo Multiplier Tube (MaPMT) readout system \cite{hamamatsu2024}, which selected the emission characteristics of light emitted at different shower depths using Optical Filters (OPF), one per scintillator emission, placed in front of three of the four photocathodes of the PMT. The fourth photocathode was equipped with a $420$ nm bandpass filter (BPF), which was chosen to avoid interference with the emission characteristics of the other scintillator materials.

We evaluated the photon yield across all output channels through test beam data analysis. These evaluations provide critical insights into the chromatic calorimeter's performance and the dynamics of electromagnetic showers. Our study focuses on distinguishing between electrons and pions, profiling the shower along its full length, and particle identification through pulse shape analysis. These findings underscore the importance of making precise adjustments in detector design to enhance the accuracy of chromatic reconstruction.

\subsection{Test Beam Setup}
\begin{figure}[H]
\centering
\includegraphics[width=12cm,clip]{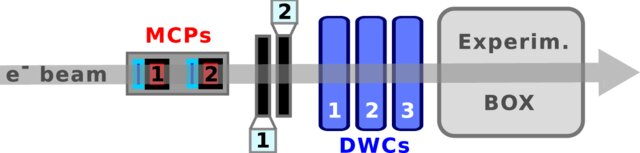}
\caption{Test beam configuration: The electron beam travels from left to right. Two MCPs (Micro-Channel Plates) establish the time reference, while two scintillating pads generate the trigger signal. Three Drift Wire Chambers (DWC) supply the tracking information used for the module alignment. The experimental box houses our prototype and the rotating stepper motors (from \cite{AN2023167629}).}
\label{fig-1ii}       
\end{figure}

The tested chromatic calorimeter module was placed in the experimental box and exposed to a beam of electrons or pions of up to 100 GeV, (see Fig. \ref{fig-1ii}). The test beam experiment was performed at the Super Proton Synchrotron (SPS) facility in the zone H2/H6 Northern area of CERN in July 2024 \cite{cern2024}.

\appendix
\section{Results}
This section presents a comprehensive analysis of the tested calorimeter module's performance, beginning with the test beam analysis to study the longitudinal shower profile. This profile provides valuable insights into the chromatic separation between output channels, which is crucial for understanding the longitudinal segmentation of the calorimeter. The investigation then examines shower progression and energy deposition by analyzing the amplitude fraction across the detector. It covers crucial aspects of calorimeter performance, including shower profiles, energy deposition, particle identification, and overall resolution.
\subsection{Study of longitudinal shower profile}
\begin{figure}[h]
\centering
\includegraphics[width=0.7\textwidth,clip]{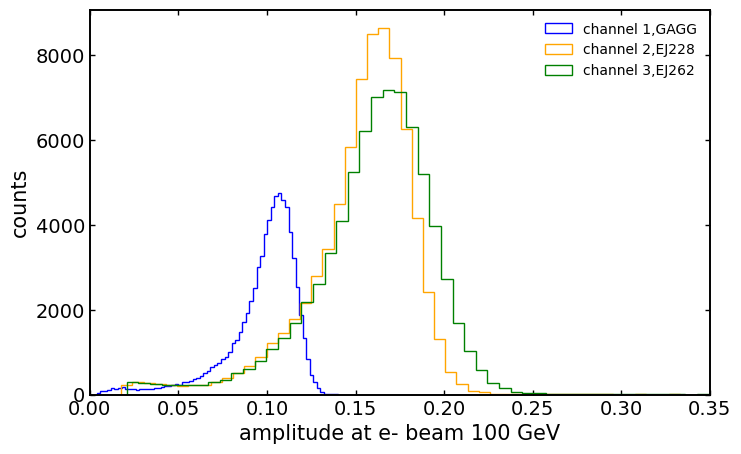}
\caption{Plot illustrating the output signal amplitude(in arb. units) measured by the three output channels of MaPMT for 100 GeV electrons, where channels 1, 2, and 3 correspond to the average signal amplitude response from GAGG, EJ228, and EJ262 respectively. The output signal amplitude is normalized w.r.t the LY values of the scintillating materials.}
\label{fig-2i}       
\end{figure}

\begin{figure}[H]
\centering
\includegraphics[width=0.7\textwidth,clip]{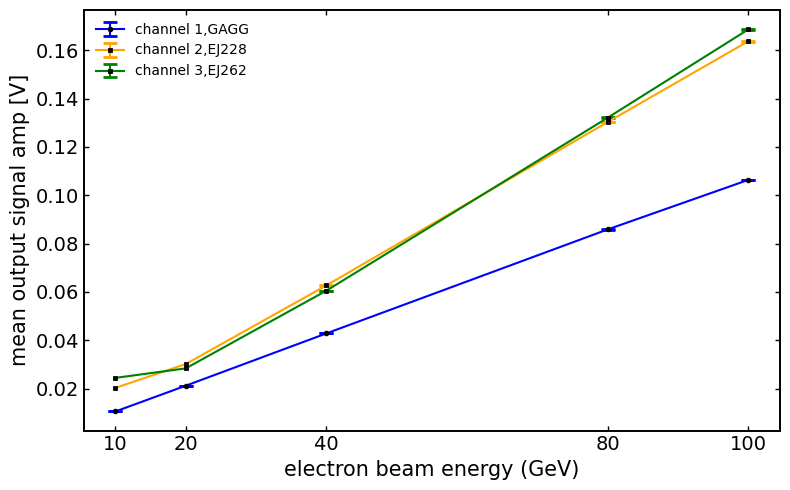}
\caption{Plot illustrating the mean of the output signal amplitude with different electron beam energies of 3 output channels, where channels 1, 2, and 3, correspond to the average signal amplitude response from GAGG, EJ228, and EJ262. The curve is the result of the crystal-ball fit equations. The output signal amplitude is normalized w.r.t the LY values of the scintillating materials.}
\label{fig-2ii}       
\end{figure}

In the analysis of the test beam, we measured the output signal amplitudes from various channels of the MaPMT at electron beam energies of 10, 20, 40, 80, and 100 GeV. Channels 1, 2, 3, and 4 were tuned to correspond with the peak emissions from GAGG, EJ228, EJ262, and a 420 nm band-pass filter using optical filters. The amplitude spectra recorded for the 100 GeV electron beam are presented in Fig. \ref{fig-2i}. This plot demonstrates a distinct chromatic separation among the channels, which becomes more pronounced as the energy of the electron beam increases. This separation aids in the accurate identification of the energy levels of incoming particles. In Fig. \ref{fig-2ii}, we show the average output amplitude for channels 1, 2, and 3 across the various energy levels (10, 20, 40, 80, and 100 GeV). The averages were calculated by fitting the channel amplitudes with a Crystal Ball function \cite{Gaiser:1982yw}, which accounts for the exponential tails. By averaging the signals from these channels, we can observe their responses to various energy levels, and the differences in average amplitudes further emphasize the concept of chromatic separation.
The output signal amplitude is normalized concerning the scintillating materials' respective light yield (LY) values as depicted in Table 1.

\begin{table}[h]
\centering
\begin{tabular}{|l|c|}
\hline
\textbf{Stack Materials} & \textbf{LY (photons/MeV)} \\
\hline
GAGG \cite{gaggly} & 60,000 \\
PbF\textsubscript{2} \cite{pbf2ly} & 0.38 \\
EJ-262 \cite{scionixnl} & 8,700 \\
EJ-228 \cite{scionixnl} & 10,200 \\
\hline
\end{tabular}
\caption{LY of Scintillator materials in the Chromatic calorimeter stack}
\label{tab:light_yield}
\end{table}

The overlap between EJ228 and EJ262 signals indicates that the optical filtering system doesn't completely separate their emissions, despite using different optical filters. However, GAGG's signal is well-isolated, showing the effectiveness of the long-pass filter configuration for this channel. Notably, we observe contamination in the output, indicating overlapping contributions from different output channels.
The amplitude distributions reveal distinct signal patterns for three scintillator channels responding to a 100 GeV electron beam, with GAGG's signal well-separated at lower amplitudes (peaking around 0.1) while EJ228 and EJ262 show significant overlap at higher amplitudes (peaking around 0.17-0.18). The similar amplitudes of the plastic scintillators, despite their different positions in the stack, are particularly noteworthy. While optical filtering effects contribute to this overlap, the normalization concerning light yield values (EJ228: 10,200 photons/MeV vs EJ262: 8,700 photons/MeV) may also be influencing the apparent similarity in their signal amplitudes, potentially masking the expected differences in energy deposition between the front and rear sections of the stack \cite{scionixnl}.


\subsection{Shower Progression and Amplitude Fraction}
\begin{figure}[h]
\centering
\includegraphics[width=0.7\textwidth,clip]{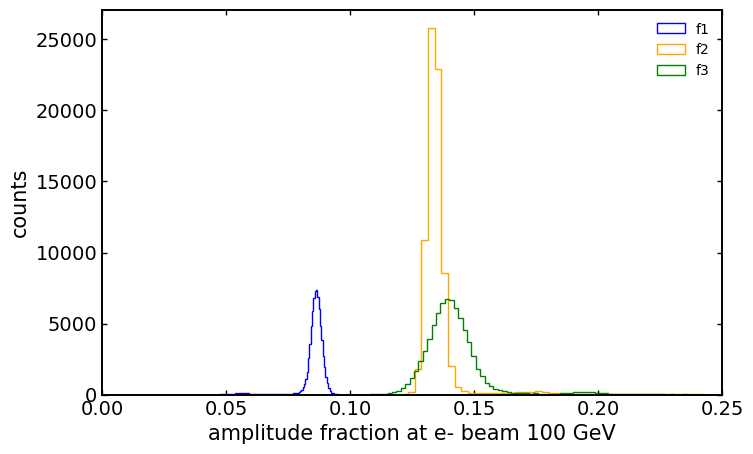}
\caption{Plot illustrating the output signal amplitude fraction measured by the three output channels of MaPMT for 100 GeV electrons, \(f_1\), \(f_2\), and \(f_3\) correspond to the average signal amplitude fraction response from GAGG, EJ228, and EJ262, respectively. The output signal amplitude is normalized w.r.t the LY values of the scintillating materials.}
\label{fig-2iii}       
\end{figure}

\begin{figure}[H]
\centering
\includegraphics[width=0.7\textwidth,clip]{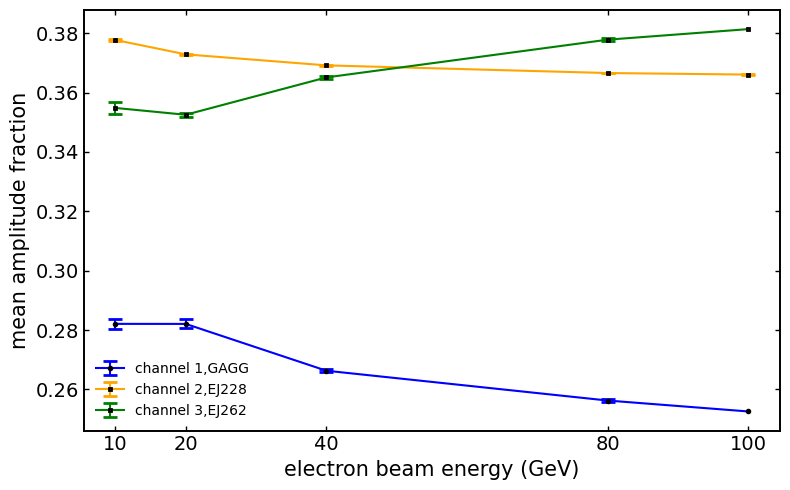}
\caption{Plot illustrating the mean of the output signal amplitude fraction with different electron beam energy of 3 output channels, where channels 1, 2, and 3 correspond to the output signal amplitude fraction of GAGG, EJ228, and EJ262. The curve is the result of the crystal-ball fit equations. The output signal amplitude is normalized w.r.t the LY values of the scintillating materials.}
\label{fig-2iv}       
\end{figure}

To study the energy distribution across the different channels of the chromatic calorimeter, we introduce the amplitude fraction as a metric to quantify the relative contribution of each output channel to the total signal. The amplitude fraction for a given channel \(f_i\) is defined as:

\[
f_i = \frac{A_i}{\sum\limits_{j} A_j}
\]
where $A_i$ is the detected output amplitude from the $i^{th}$ channel.
At an electron beam energy of 100 GeV, the plot of amplitude fraction \(f_1, f_2, f_3\) as a function of the shower progression provides a clear view of how the energy deposition is distributed across the different regions of the detector, as shown in Figure \ref{fig-2iii}. Each fraction describes how much of the total signal is captured by a particular channel, offering insights into the longitudinal development of the shower and the contribution of each scintillator material.
\\
\textbf{Amplitude Fraction Interpretation:} At 100 GeV electron beam energy, the amplitude fraction distributions show distinct characteristics across the three channels, as shown in Figure \ref{fig-2iii}. The amplitude fraction \(f_1\) exhibits a narrow peak centered at approximately 0.09, while \(f_2\) displays a sharp, prominent peak at around 0.13-0.14. The amplitude fraction \(f_3\)  shows a broader distribution centered near 0.15. The significant overlap between \(f_2\)  and \(f_3\)  in the 0.13-0.15 region indicates signal contamination between output channels, while \(f_1\) maintains a well-separated distribution at lower amplitude fractions. This distribution pattern reflects the energy deposition behavior across the calorimeter, with \(f_3\) showing increased energy deposition at higher beam energies compared to the other channels.
\\
\textbf{Mean Amplitude Fraction Across Different Beam Energies}:
We calculated the average amplitude fraction across different electron beam energies and analyzed it at a single beam energy. This is done by averaging the amplitude fractions \(f_1, f_2, f_3, f_4\) at various beam energies, ranging from lower to higher energy levels.
The plot of the mean amplitude fraction as a function of electron beam energy provides insight into how the energy deposition pattern evolves with increasing energy, see Figure \ref{fig-2iv}. As the energy of the incident particle increases, the longitudinal spread of the shower broadens, which is reflected in the shifting amplitude fractions of the channels. The variation in mean amplitude fraction at higher beam energies is consistent with the depth of the shower maximum that shifts as the incident energy increases. Channel 1's decreasing trend indicates that more energy is deposited in the downstream channels as the shower progresses deeper into the calorimeter. This is evident by channel 3's increasing amplitude fraction, particularly after 40 GeV where it crosses channel 2's stable response, see Figure \ref{fig-2iv}. We also observe contamination in the output, indicating overlapping contributions from different output channels. This is particularly evident in the crossing point between channel 2 and channel 3 at approximately 40 GeV, where their signals begin to overlap, suggesting that the energy deposition patterns become less distinct at higher energies.
This behavior demonstrates the longitudinal shower development, with more energy being deposited in later layers (EJ262) as beam energy increases while maintaining stable fractions in front sections (GAGG and EJ228).
This analysis is crucial for understanding how the calorimeter performs across a range of beam energies and optimizing its design for future experiments.
\subsection{Electron-Pion Separation}
\begin{figure}[ht]
\centering
\includegraphics[width=\textwidth,clip]{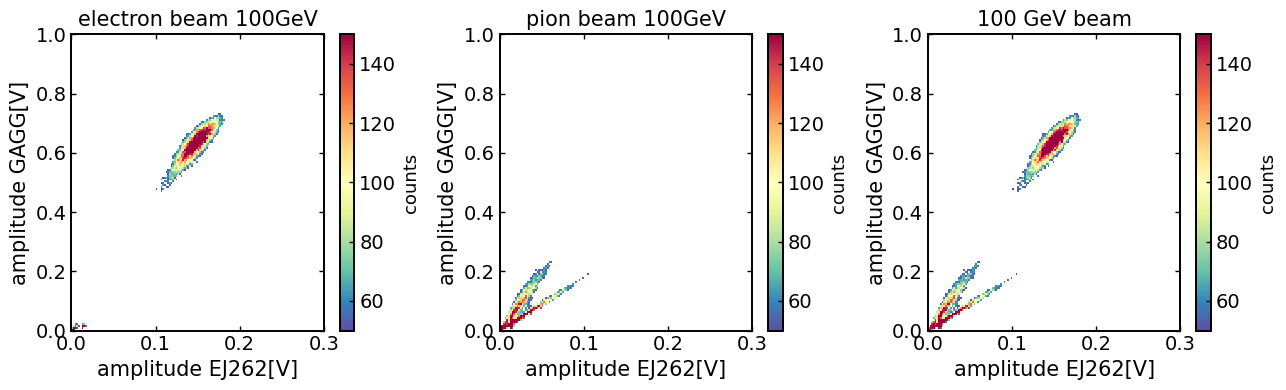}
\caption{Scatter plots display the correlation between signal amplitudes recorded in GAGG and EJ262 for electrons (e$^-$) and charged pions ($\pi^+$) at 100 GeV. Each point corresponds to an event, with the x-axis representing the signal amplitude in GAGG and the y-axis representing the signal amplitude in EJ262. A threshold of $50$ counts is imposed.}
\label{fig-2}       
\end{figure}

It was observed that exposing the scintillator stack to high-energy electrons and pions at the (SPS) facility in the zone H2/H6 Northern area of CERN, with energies up to 100 GeV, resulted in a shift in the amplitude ratio between the responses of GAGG, EJ262, and EJ228. This shift allowed us to distinguish electrons from pions, as shown in Figure \ref{fig-2}. A comparable amplitude scatter plot can also be generated for the same particle type at different energy levels.
\\
\subsection{Center of Gravity (\( z_{\text{cog}} \)) and Shower Depth:}
\begin{figure}[H]
\centering
\includegraphics[width=0.7\textwidth,clip]{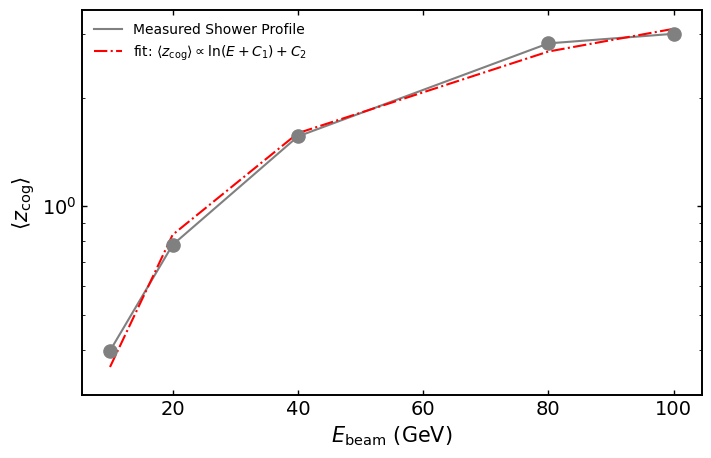}
\caption{Evolution of the center of gravity \( \langle z_{\text{cog}} \rangle \) as a function of beam energy in the chromatic calorimeter. The grey line represents the measured shower profile, while the red dashed line shows the logarithmic fit \( \langle z_{\text{cog}} \rangle \propto \ln(E + C_1) +C_2 \) Ref\cite{Workman:2836514}}
\label{fig-2v}       
\end{figure}

The center of gravity \( \langle z_{\text{cog}} \rangle \) is a useful metric for describing the average depth of energy deposition in calorimeters during electromagnetic showers \cite{Bonanomi:2797465}. The shower progression in the chromatic calorimeter, as represented in Figure \ref{fig-2v}, shows the evolution of the center of gravity \( \langle z_{\text{cog}} \rangle \) with increasing beam energy. The plot highlights the gradual shift of energy deposition into the calorimeter as the beam energy increases, aligning with the expected logarithmic growth. This is evidenced by the clear trend in \( \langle z_{\text{cog}} \rangle \) versus energy, where higher energies correspond to deeper penetration into the detector layers. The radiation lengths of the materials are taken into account for this calculation \cite{Gupta:1279627}. The fit applied to the data supports this observation, illustrating a relationship of \( \langle z_{\text{cog}} \rangle \propto \ln(E+C_1) + C_2 \), indicating that the depth of energy deposition grows logarithmically with electron beam energy. This means that if we increase the energy of the beam, the average depth where the energy is deposited in the chromatic calorimeter increases at a decreasing rate. In simpler terms, the first few increases in energy lead to large increases in depth, but as we keep increasing the energy, the additional depth gained becomes smaller. The constants $C_1$ and $C_2$ in the equation, help to adjust the curve to fit the test beam data accurately. $C_1$ shifts the logarithm curve to the right, helping to account for the initial energy levels, where the shower starts. At the same time, $C_2$ adjusts the vertical position of the curve, which ensures the average depth of the energy deposition observed. The logarithm relationship suggests that the relationship is not linear, which is consistent with understanding how particles interact with matter. \\
\subsection{Calibration}In this section, we describe the calibration process for the chromatic calorimeter using test beam data from Minimum Ionizing Particles (MIPs), i.e. muons at various energy levels. We implemented muons as the reference channel to ensure consistent energy deposition readings across the MaPMT due to their stable energy characteristics at varying pion beam energies. In amplitude fraction ($f_i$) scatter plots (refer to Section $A.2$), the muons were identified as a cluster of detected particles consistent with increasing pion beam energy.
Hence, we introduce a calibration approach known as the fraction method by MIPs, which utilizes data from pion beam runs to establish a correction factor, denoted as \( K_i(E_\text{beam})\). This correction factor quantifies the relationship between the amplitude of the reference channel \( A_0(E_\text{beam}) \) and that of $i^{th}$ channel \( A_i(E_\text{beam}) \), allowing us to adjust their responses accordingly.
\[
K_i(E_\text{beam}) = \frac{A_0(E_\text{beam})}{A_i(E_\text{beam})}
\]
Using this correction factor, the response for each layer of the calorimeter is expressed as:
\[
S_i(E_\text{beam}) = K_i(E_\text{beam}) A_i(E_\text{beam})
\]
where \( E_\text{beam} \) is the incoming beam energy, \( A_0(E_\text{beam}) \) is the amplitude response from MIPs and \( A_i(E_\text{beam}) \) is the measured response from the MaPMT in $i^{th}$ channel.\\ 
Future studies will refine this correction factor for photo-detector efficiencies, reflections, and potential signal attenuation variations.
\subsection{Energy reconstruction and resolution}
Energy reconstruction and resolution are crucial aspects of measuring the energy of particles in experiments. Energy reconstruction refers to estimating the energy of an incoming particle based on the signals recorded by the detector. A good reconstruction method ensures that the estimated energy closely matches the actual energy of the particle \cite{simon2009energy}. On the other hand, energy resolution describes how accurately we can measure that energy. It indicates the uncertainty in our energy measurements—lower values signify better precision.
\begin{figure}[ht]
    \centering
    \begin{subfigure}{0.49\textwidth}
        \centering
        \includegraphics[width=\textwidth,clip]{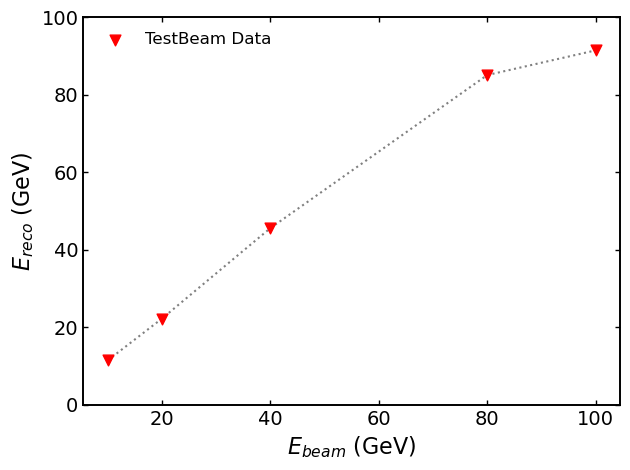}
        \caption{}
        \label{fig-2vi}
    \end{subfigure}
    \hfill
    \begin{subfigure}{0.49\textwidth}
        \centering
        \includegraphics[width=\textwidth,clip]{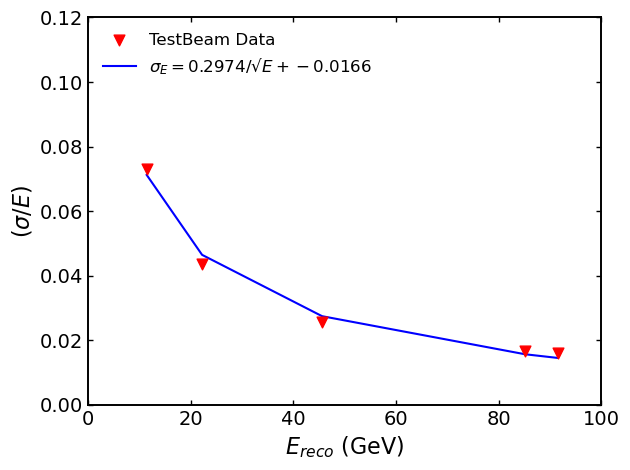}
        \caption{}
        \label{fig-2vii}
    \end{subfigure}
    \caption{Plots (a) and (b), illustrate the energy reconstruction and resolution in the chromatic calorimeter stack, respectively.}
    \label{fig:energy_reconstruction}
\end{figure}
Fig \ref{fig-2vi}, displays the relationship between the reconstructed and actual incident electron energies. As shown, the reconstructed energy values increase as the incident electron energy rises, indicating that our energy reconstruction method is effective. This plot also shows that for electron energies up to 100 GeV, the reconstructed values align well with the expected energies.

Figure \ref{fig-2vii}, illustrates how energy resolution (expressed as $\sigma/E$, where $\sigma$ is the uncertainty in the energy measurement) varies with the incident electron energies. The resolution values range from about $0.08$ at lower energies to approximately $0.02$ at higher energies, demonstrating that our measurements become more precise as the energy of the incoming electrons increases.

These plots highlight the success of our chromatic calorimetry experiment in accurately reconstructing energy and improving measurement precision across a range of incident energies.
Together, these plots illustrate the effectiveness of our energy reconstruction technique and highlight improvements in measurement accuracy across different energy levels. We analyzed our experimental method’s energy reconstruction and resolution by employing suitable techniques. The energy reconstruction is performed using the formula:
\[
E_{\text{reco}} = \sum_{i} c_i A_i (E_{\text{beam}})
\]
where \( E_{\text{reco}} \) represents the reconstructed energy, \( c_i \) are the coefficients to be determined, and \( A_i (E_{\text{beam}}) \) are the amplitudes corresponding to the beam energy. To optimize the coefficients \( c_i \), we minimize the chi-squared statistic defined as:
\[
\chi^2 = (E_{\text{reco}} - E_{\text{beam}})^2
\]
This approach allows us to assess the energy resolution accurately and ensure that our reconstruction aligns closely with the expected beam energy.

\section{Discussion, conclusion, and outlook}
We present the findings from the test beam analysis of the modified experimental iteration of our chromatic calorimetry project. Our study clearly distinguishes between electron and pion events, demonstrating that the reconstructed energy signals effectively differentiate between hadronic and electromagnetic interactions, particularly at higher energy levels. This capability allows for improved event classification, as the signals show a consistent progression concerning the energy of incoming particles, indicating enhanced resolution and sensitivity across various energy ranges. These results highlight the effectiveness of our reconstruction method and its potential for accurate particle identification. Notably, we also observe contamination in the output, indicating overlapping contributions from different output channels. This observation underscores the need for further improvements in our experimental setup. We plan to conduct additional analysis and simulations based on our observations. We aim to refine our next experimental setup by optimizing filters, testing new materials, and exploring other enhancements. While chromatic calorimetry utilizes varying scintillator materials to differentiate emitted light wavelengths and introduces innovative approaches to longitudinal segmentation, it is important to note that our current experiments do not incorporate quantum dots. Instead, we focus on optimizing the arrangement of scintillator materials to maximize performance in particle detection.
\acknowledgments
This work was carried out as part of the CERN Quantum Technology Initiative (QTI), the CERN Crystal Clear Collaboration (CCC), and the ECFA-DRD5 collaboration.

\section*{Conflict of Interest}
The authors declare that the research was carried out without commercial or financial ties that might be considered a conflict of interest.\\




\bibliographystyle{JHEP}
\bibliography{biblio} 

\end{document}